\documentclass[proceedings]{rmaa}

\renewcommand{\P}[1]{%
\ifnum#1=1\hbox{OW~168--326E}\fi
\ifnum#1=2\hbox{OW~167--317}\fi
\ifnum#1=3\hbox{OW~163--317}\fi
\ifnum#1=5\hbox{OW~158--323}\fi
\ifnum#1=0\hbox{OW~171--334}\fi}

\title { Ultra-High-Energy Cosmic Ray Acceleration by Magnetic 
Reconnection in Newborn  Pulsars}
\author{Elisabete M. de Gouveia Dal Pino\altaffilmark{1,2}
\& Alex Lazarian\altaffilmark{3} } 
\altaffiltext{1}{Instituto Astron\^omico e Geof\'{\i}sico, University 
of S\~ao Paulo, E-mail: dalpino@iagusp.usp.br } 
\altaffiltext{2}{
Astronomy Department,
University of California at Berkeley} 
\altaffiltext{3}{Department of  Astronomy, 
University of Wisconsin }

\fulladdresses{
\item E. M. de Gouveia Dal Pino: 
IAG-USP, Av. Miguel St\'efano, 4200, S\~ao Paulo
04301-904, SP, Brasil} 

\shortauthor{de Gouveia Dal Pino \& Lazarian}
\shorttitle{UHECR Acceleration by Reconnection}


\abstract{  We here investigate the possibility that the 
ultra-high energy  cosmic ray (UHECR) events observed above the GZK 
limit are mostly protons accelerated in reconnection sites just above 
the 
magnetosphere of newborn millisecond pulsars 
that are originated by accretion 
induced collapse (AIC).
  }

\nonstopmode

\begin{document}

\maketitle

\section{Introduction}
\label{sec:intro}

The detection of  cosmic ray events with energies beyond 10$^{20}$ eV
(UHECRs) poses a challenge for the understanding of their 
nature
and sources.  If UHECRs are mostly
protons, then they should be 
affected 
by
the expected Greisen-Zatsepin-Kuzmin (GZK) energy cutoff 
($\sim 5\times
10^{19}$ eV), which is due to photomeson production by interactions 
with 
the
cosmic microwave background radiation, unless they are originated at
distances closer than about 50 Mpc  (e.g., 
Medina Tanco, de Gouveia Dal Pino \& Horvath 1997). On the other hand, 
if
the UHECRs are mostly protons from nearby sources 
(located within $\sim $
 50 Mpc), then the arrival directions of the events should point toward
their sources since they are expected to be little deflected by the
intergalactic and Galactic magnetic fields (e.g., Medina 
Tanco, de Gouveia Dal Pino \& Horvath 1998). The present data shows no 
significant
large-scale anisotropy in the distribution related to the Galactic disk 
or
the local distribution of galaxies, although some clusters of events 
seem 
to
point to the supergalactic plane (Takeda et al. 1999).

We here discuss a model in which UHECRs are mostly protons 
accelerated in
magnetic reconnection sites outside the magnetosphere of  very young 
millisecond pulsars being produced by accretion induced collapse (AIC) 
of a 
white dwarf (de Gouveia Dal Pino \& Lazarian 2000; hereafter GL2000). 
When a white dwarf reaches the critical Chandrasekhar mass $\sim 1.4$
M$_{\odot }$ through mass accretion, in some cases it collapses directly to
 a neutron star instead of exploding
into
a supernova. The accretion flow spins up the star and confines the 
magnetosphere  
to a radius $R_X$ where  plasma stress in the accretion disk  and 
magnetic 
stress 
balance
(Arons 1993). At this radius the equatorial flow will divert into a
funnel inflow
along the closed 
field-lines toward the star, and 
a centrifugally 
driven wind outflow  (see Fig. 1 of GL2000). 
To mediate the  field lines of the star with 
those
opened by the wind and those trapped by the funnel inflow emanating 
from
the $R_X$ region a surface of null poloidal field forms 
(e.g., Shu et al. 1994). 
This reconnection region dominated "$helmet$ $streamer$",
will release  magnetic energy that will accelerate particles to 
the UHEs.

A primary condition on the reconnection region 
 for it to be
able to accelerate particles of charge $Ze$ to energies $E$ 
 is that its  width
$\Delta R_X \,  \geq \, 2 \, r_L$, 
where 
$r_L$ is the particle Larmour radius 
$r_L = E /Z e \, B_X$ , and 
$B_X$  is the magnetic 
field (normal to particle velocity) at the $R_X$ region. 
This condition and the field geometry imply (GL2000):
\begin{equation}
B_{13} \, \gtrsim \, Z^{-1} \,  E_{20} \, \Omega_{2.5k}^{-4/3} \, 
   \left({\frac{\Delta R_X/R_X } { 0.1}}\right)^{-1/2} 
\end{equation}
\noindent 
where $B_{13}$ is the stellar magnetic field in units of 
$10^{13}$ G, 
$\Omega_{2.5k}$ is the stellar angular speed in units of 2500 s$^{-1}$,
and   $E_{20}$ is the particle energy in units of 10$^{20}$ eV.
 We find that  stellar magnetic fields 
 $10^{12} $ G $ <  B_{\star} \lesssim   10^{15}$ G 
and angular speeds 
$4 \times 10^{3}$ s$^{-1}$ $\gtrsim \Omega_{\star} \, > \,  10^{2} $ 
s$^{-1}$
(or spin periods
1 ms $\lesssim \, P_{\star} \, < \, $ 60  ms), are able to accelerate
particles to energies $E_{20} \,  \gtrsim $ 1.

A newborn millisecond pulsar spins down due to 
magnetic 
dipole radiation in a time scale given by
$\tau_{\star} \equiv \Omega_{\star}/\dot \Omega_{\star}
\simeq 4.3 \times 10^7 $ s $  B_{13}^{-2} \, 
\Omega_{2.5k}^{-2}$. 
We can show that the 
condition that the magnetosphere and disk stresses are in equilibrium 
at 
the inner disk edge results a disk mass accretion rate 
that is 
 super-Eddington. This supercritical accretion  will  
 last for $\sim $ $\tau_{\star}$. 
 As it approaches the 
end, the 
newborn pulsar decreases its rotation speed due to electromagnetic 
radiation at a rate $\tau_{\star}^{-1}$. 
The spectrum evolution of 
the 
accelerated UHECRs is thus  determined by 
$\tau_{\star}^{-1}$. 
The particle spectrum $N(E)$ is obtained from
$\dot N \, = \, N(E) \, {\frac {dE } { dt}}  \, 
= \, N(E) \,  {\frac {dE } { d \Omega_{\star} }  } \, \dot 
\Omega_{\star}$
 (GL2000):
\begin{equation}
N(E) \, \simeq 
\, 5.8 \times 10^{34} \, {\rm GeV}^{-1} \,  \xi \, Z^{-1/2} \,  
B_{13}^{-
1/2} \, E_{20}^{-3/2} \, \left({\frac{\Delta R_X/R_X } 
{0.1}}\right)^{-1/4}
\end{equation}
\noindent 
where $\xi$ is the reconnection efficiency factor; 
the derived spectrum above is
 very flat which is in 
agreement with the observations.

The total number of objects formed via AICs 
in our Galaxy is limited by nucleosynthesis constraints to a very small 
rate  
$ \sim \, 10^{-5} $ yr$^{-1}$. 
Hence, the probability of having UHECR events produced in the Galaxy 
will be only
$P \, \simeq \, f_b \, {\tau_{AIC}}^{-1}  \, t  \, \simeq \, 2 \times  
10^{-
6}$, where $f_b \sim \,  (\Delta R_X/R_X)^2 \simeq 10^{-2}$ is
 the emission beaming factor caused by the magnetic field geometry, and
$t = 20 $
yr accounts for the time the UHECR events have been collected  in 
Earth detectors so far. 
Since the individual contribution to the observed UHECRs due to  
AICs in our  Galaxy is so small we must evaluate the integrated 
contribution due to  AICs from all the galaxies located within a
volume 
which is not affected by the GZK effect, i.e., within a radius 
$R_{50} = R_G/ 50$ Mpc.
Assuming  that each galaxy has essentially the
same rate of AICs as our Galaxy and taking the standard galaxy 
distribution 
$n_G \simeq \, 0.01 \, e^{\pm 0.4}\, h^3 $ Mpc$^{-3}$ 
(with the Hubble parameter defined as $H_o = h$ 100 km s$^{-1} $ 
Mpc$^{-1}$), the resulting flux at $E_{20} \, \geq $  1 is
$F(E) \, \simeq \, \,  N(E) \, n_G \, {\tau_{AIC}}^{-1} \, R_{G}$, 
which gives
\begin{equation}
F(E) \, \simeq \, 1.1 \times 10^{-27} \xi \, {\rm GeV}^{-1} 
{\rm cm}^{-2} {\rm s}^{-1} \,
Z^{-1/2} \, B_{13}^{-1/2} \, 
E_{20}^{-3/2} \, {\tau_{AIC,5}}^{-1} \, n_{0.01}  \, 
R_{50} \left({\frac{\Delta R_X/R_X } {0.1}}\right)^{-1/4} 
\end{equation}
\noindent 
where ${\tau_{AIC,5}}^{-1} \, = \, {\tau_{AIC}}^{-1}/ 10^{-5}$ yr$^{-
1}$,
 and  
$n_{0.01} = n_G/0.01 $ h$^3$ Mpc$^{-3}$.
Observed data by the AGASA experiment (Takeda et al. 1999) gives a flux 
at 
$E = $10$^{20}$  eV  of 
$F(E) \, \simeq  \,  4 \times \, 10^{-30}$ Gev$^{-1}$ cm$^{-2}$ s$^{-
1}$, so 
that the reconnection efficiency factor needs to be only
$ \xi \, \gtrsim \, 3.6 \times 10^{-3}$
in order to reproduce the observed signal.


\acknowledgements E.M.G.D.P. has been partially supported by a grant of the Brazilian 
Agency FAPESP and by the PRONEX. 


\end{document}